\documentclass[aps,pra,twocolumn]{revtex4}
\usepackage{graphicx}
\usepackage{dcolumn}
\usepackage{amsmath}
\usepackage{amssymb}
\def\erf{{\rm erf}}
\def\rv{{\bf r}}

\def\Rv{{\bf R}}

\def\beq{\begin{equation}}
\def\eeq{\end{equation}}

\begin{document}
\title{System-adapted correlation energy density functionals from
effective pair interactions}
\author{Paola Gori-Giorgi and Andreas Savin}
\affiliation{Laboratoire de Chimie Th\'eorique, CNRS,
Universit\'e Pierre et Marie Curie, 4 Place Jussieu,
F-75252 Paris, France}
\date{\today}
\begin{abstract}
We present and discuss some ideas
concerning an
``average-pair-density functional theory'', 
in which the ground-state energy of
a many-electron system is rewritten as a functional of the
spherically and system-averaged pair density $f(r_{12})$. 
These ideas are further clarified 
with simple  physical examples. We then show
that the proposed formalism can be combined with density functional
theory to build system-adapted correlation energy functionals.
A simple approximation for the unknown effective electron-electron
interaction that enters in this combined approach is described, and
results for the He series and for the uniform electron gas
are briefly reviewed. 
\end{abstract}

\maketitle
\section{Introduction}

Density Functional Theory (DFT) is nowadays 
the most widely used method for the calculation
of electronic structure in both solid-state physics and
quantum chemistry~\cite{kohnnobel,science,FNM}. The accuracy
of the results coming from a DFT calculation is limited by
the approximate nature of the exchange and correlation
energy $E_{xc}[n]$ and the associated potential, its functional
derivative.
While, at present, many algorithms for a very accurate
or even exact exchange potential $v_x(\rv)$ and energy $E_x[n]$
are available~\cite{FNM,oep}, better approximations for calculating accurate
correlation energies $E_c[n]$ are still needed.

In a recent paper~\cite{GS1}, we proposed an alternative approach to
build the DFT correlation energy $E_c[n]$. The method consists in solving
simple radial equations to generate the spherically and system-averaged
pair density $f(r_{12})$ along the so-called adiabatic connection.
Besides its practical use for DFT, we realized that 
this method has many aspects that deserve to be 
better investigated, including the possibility
for an alternative theory
completely based on $f(r_{12})$. In this work, we sketch the
basic ideas for this alternative theory, we further discuss them
with simple physical examples, and we deal
with some of the aspects that went overlooked in Ref.~\cite{GS1}.
The scope of this paper is to lay the rigorous foundations
for an approach that will be further developed towards the construction
of a self-consistent scheme combining the radial equations for
$f(r_{12})$ with the Kohn-Sham equations.

The paper is organized as follows. After defining the notation,
we develop in Sec.~\ref{sec_formalism}  the formalism corresponding
to a theory based on the spherically and system averaged
pair density, reviewing at the same time the corresponding
concepts for DFT. We hope that this parallel treatment will make it easier 
for the reader to familiarize with the new concepts.
We then give, in Sec.~\ref{sec_ex}, some simple physical examples. 
Section~\ref{sec_comb} is devoted to explain how the ideas
of Sec.~\ref{sec_formalism} can be used to build correlation energy functionals
for DFT. A simple, physically motivated, approximation for the
unknown effective eletron-electron interaction that appears in our
formalism is discussed in Sec.~\ref{sec_app}, where
 applications to two-electron atoms and to the uniform electron gas
are also briefly reviewed. The last section is devoted to conclusions
and perspectives.
  
\section{Definitions}
\label{sec_def}
We start from the standard $N$-electron hamiltonian (in Hartree atomic units,
$\hbar=m=a_0=e=1$, used throughout)
\begin{eqnarray}
H & =& T+V_{ee}+V_{ne}, \\
T & = & -\frac{1}{2}\sum_{i=1}^N\nabla_i^2, \\
V_{ee} & = & \frac{1}{2}\sum_{i\neq j}^N \frac{1}{|\rv_i-\rv_j|}, \\
V_{ne} & = & \sum_{i=1}^N v_{ne}(\rv_i),
\end{eqnarray}
where $v_{ne}$ is the external potential due to nuclei.
Given $\Psi$, the ground-state wavefunction of $H$, we consider
two reduced quantities that fully determine, respectively,
 the expectation values
$\langle\Psi|V_{ne}|\Psi\rangle$ and $\langle\Psi|V_{ee}|\Psi\rangle$, i.e.,
the usual one-electron density 
\beq
n(\rv)=N\sum_{\sigma_1...\sigma_N}\int |\Psi(\rv,\rv_2,...,
\rv_N)|^2d\rv_2...d\rv_N,
\eeq
and the spherically and system-averaged pair density (APD),
which is obtained as an integral of $|\Psi|^2$ over all variables but
$r_{12}=|\rv_1-\rv_2|$,
\begin{eqnarray}
f(r_{12}) = \frac{N(N-1)}{2}\sum_{\sigma_1...\sigma_N} \times \nonumber \\ 
 \int |\Psi(\rv_{12},\Rv,\rv_3,...,\rv_N)|^2
\frac{d\Omega_{\rv_{12}}}{4\pi} d\Rv d\rv_3...d\rv_N,
\end{eqnarray}
where $\rv_{12}=\rv_1-\rv_2$, and $\Rv=\tfrac{1}{2}(\rv_1+\rv_2)$.
The function $f(r_{12})$ is also known in chemistry as intracule
density~\cite{thakkar,Coulson,Cioslowski1,Cioslowski2,ugalde1,davidson,coleman},
and, when multiplied by the volume element $4\pi r_{12}^2dr_{12}$, is 
proportional to the probability distribution for the electron-electron
distance.
We then have
\begin{eqnarray}
\langle\Psi|V_{ne}|\Psi\rangle  =  \int n(\rv) v_{ne}(\rv) d\rv \\
\langle\Psi|V_{ee}|\Psi\rangle  =  \int \frac{f(r_{12})}{r_{12}} d\rv_{12}=
\int_0^{\infty}\frac{f(r_{12})}{r_{12}}4\pi r_{12}^2 dr_{12}.
\end{eqnarray}
In the following text we will also deal with modified systems in which 
the external potential
and/or the electron-electron interaction is changed. Thus, 
the notation $V_{ee}$ and $V_{ne}$ will
be used to characterize the physical system, while the modified systems will be defined by
$W$ and $V$, with
\begin{eqnarray}
W & = & \frac{1}{2}\sum_{i\neq j}^N w(|\rv_i-\rv_j|), \\
V & = & \sum_{i=1}^N v(\rv_i),
\end{eqnarray}
where the pairwise interaction $w$ always depends only on
$|\rv_i-\rv_j|$.

\section{Formalism}
\label{sec_formalism}
In this section we present a ``APD-functional
theory'' (APDFT) based on the function $f(r_{12})$ highlighting, step
by step, the analogies in reasoning with the derivation of standard DFT. 

\subsection{DFT -- The universal functional}
In standard DFT one defines a universal functional of the one-electron density
$n$ as resulting from a constrained search over all the antisymmetric
wavefunctions $\Psi$ that yield $n$~\cite{levy}
\beq
\tilde{F}[n;V_{ee},T]=\min_{\Psi \to n} \langle \Psi |T+V_{ee}|\Psi\rangle,
\label{eq_dftlevy}
\eeq
or, more completely, as a Legendre transform~\cite{lieb}
\begin{eqnarray}
F[n;V_{ee},T]& = &\sup_v\Big\{\min_{\Psi} \langle \Psi |T+V_{ee}+V|\Psi\rangle
\nonumber \\
& & -\int n(\rv) v(\rv) d\rv\Big\}.
\label{eq_dftlieb}
\end{eqnarray}
In both Eqs.~(\ref{eq_dftlevy}) and~(\ref{eq_dftlieb}), the dependence
on the electron-electron interaction (and on the kinetic energy operator $T$)
 has been explictly shown in the functional. The universality of the functional
$F$ stems exactly from the fact that the e-e interaction is always
$1/r$ (and that $T$ is always the same).

The ground-state energy $E_0$ of the system can then be obtained by minimizing 
the energy with respect to $n$,
\beq
E_0=\min_n\left\{F[n;V_{ee},T]+\int n(\rv) v_{ne}(\rv) d\rv\right\}.
\label{eq_Edft}
\eeq

\subsection{APDFT -- The system-dependent functional}
Similarly, we can define a system-dependent functional (i.e., a functional
depending on the external potential $V_{ne}$, and thus on the specific system) 
of the APD $f(r_{12})$ as
\beq
\tilde{G}[f;V_{ne},T]=\min_{\Psi \to f} \langle \Psi |T+V_{ne}|\Psi\rangle,
\label{eq_fftlevy}
\eeq
or better as
\begin{eqnarray}
G[f;V_{ne},T]& = & \sup_w\Big\{\min_{\Psi} \langle \Psi |T+W+V_{ne}|\Psi\rangle \nonumber \\
& & -\int f(r_{12}) w(r_{12}) d\rv_{12}\Big\}.
\label{eq_fftlieb}
\end{eqnarray}
The ground-state energy can be obtained by a minimization with respect
to $f$
\beq
E_0=\min_f\left\{G[f;V_{ne},T]+\int \frac{f(r_{12})}{r_{12}} d\rv_{12}\right\}.
\eeq
Evidently, with respect to DFT, the functional $G$ has the disadvantage
of being not universal: in DFT, an approximation for $F$  should be
in principle valid for all systems. However, the crucial point for
applications is understanding how
difficult is to build a reasonable approximation for $G[f;V_{ne},T]$, 
given a certain $V_{ne}$. In the particular combination of DFT and
APDFT that we propose in Sec.~\ref{sec_comb} the 
lack of universality of $G$ is not an issue.

Another issue concerns the $N$-representability conditions
on $f(r_{12})$, i.e., which contraints must a given $f$ satisfy
to guarantee that it comes from the contraction
of a $N$-electron wavefunction $\Psi$. This is a problem
shared with other generalizations of DFT, like the pair-density functional
theory~\cite{ziesche1,gonis,agnesnagy,furche,ayers}. 
 The $N$-representability conditions on $f(r_{12})$ are evidently
related to those on the pair density, and we thus might expect that 
they are not a trivial matter~\cite{ayers}. The definition
of the functional $G$ of Eq.~(\ref{eq_fftlieb}) formally overcomes this
problem by giving a divergent ($+\infty$) 
answer for any non $N$-representable $f$ (when $f$ is not 
$N$-representable the right-hand side of  Eq.~(\ref{eq_fftlieb})
is not bounded from above), but
this is not a practical solution when coming to
applications. As we shall see,
in the particular use of APDFT presented in Sec.~\ref{sec_comb}, 
this issue is not particularly crucial
 from a practical point of view, since
we use APDFT to build correlation functionals for DFT

\subsection{DFT -- Adiabatic connection}
\label{sec_DFTadia}
In density functional theory, one usually defines a set of hamiltonians
depending on a parameter $\lambda$~\cite{wang,adiabatic,gunnarsson},
\beq
H^\lambda=T+W^\lambda+V^\lambda,
\label{eq_Hlambda}
\eeq
having all the same one-electron density, equal to the one
of the physical system 
\beq
n^\lambda(\rv)=n(\rv)\qquad \forall \lambda.
\eeq
If $W^{\lambda=0}=0$ and $W^{\lambda_{\rm phys}}=V_{ee}$, one 
switches continuously
from a noninteracting system to the physical system, while keeping the
density fixed by means of a suitable external potential $V^\lambda$.
Obviously, the APD $f(r_{12})$ changes with $\lambda$.
By the Hellmann-Feynmann theorem,
\begin{eqnarray}
\frac{\partial E_0^{\lambda}}{\partial \lambda}   =  
\langle \Psi^{\lambda} |\frac{\partial W^{\lambda}}{\partial \lambda}+
\frac{\partial V^{\lambda}}{\partial \lambda}|\Psi^{\lambda}\rangle =
\nonumber \\
= \int f^{\lambda}(r_{12})\frac{\partial w^{\lambda}(r_{12})}
{\partial \lambda}d\rv_{12}+
 \int n(\rv)\frac{\partial v^{\lambda}(\rv)}
{\partial \lambda}d\rv,
\label{eq_HFDFT}
\end{eqnarray}
so that by directly integrating Eq.~(\ref{eq_HFDFT}), and by combining
it with Eq.~(\ref{eq_Edft}), one obtains
\beq
F[n;V_{ee},T]=T_s[n]+\int_0^{\lambda_{\rm phys}} d\lambda\int d\rv_{12}
f^{\lambda}(r_{12})\frac{\partial w^{\lambda}(r_{12})}
{\partial \lambda},
\label{eq_defTs}
\eeq
where $T_s[n]$ is the kinetic energy of a noninteracting system
of $N$ electrons with density $n(\rv)$.

More generally, one can be interested in using as a starting
point a system of partially interacting electrons, corresponding
to a particular value of the coupling $\lambda$ 
(say, $\lambda=\mu$) between 0 and $\lambda_{\rm phys}$. In this case,
if $\Psi^{\mu}$ is the wavefunction of the system with partial
interaction $W^{\mu}$ (and external potential $V^\mu$) we have
\begin{eqnarray}
F[n;V_{ee},T] & = & \langle \Psi^{\mu}|T+W^\mu|\Psi^{\mu}\rangle+
\nonumber \\
& & +\int_\mu^{\lambda_{\rm   phys}} d\lambda\int d\rv_{12}
f^{\lambda}(r_{12})\frac{\partial w^{\lambda}(r_{12})}
{\partial \lambda}.
\label{eq_Fmu}
\end{eqnarray}
Usually, the adiabatic connection is performed along a linear path
by setting $W^\lambda=\lambda V_{ee}$ (thus $\lambda_{\rm phys}=1$), 
but some nonlinear choices 
can be more convenient when dealing with approximations.

\subsection{APDFT -- Adiabatic connection}
\label{sec_APDFTadia}
We can also define a set of hamiltonians 
\beq
H^\xi=T+W^\xi+V^\xi,
\label{eq_Hxi}
\eeq
in which the function
$f(r_{12})$ is kept fixed, equal to the one of the physical
system,
\beq
f^\xi(r_{12})=f(r_{12})\qquad \forall \xi,
\eeq
If $V^{\xi=0}=0$ and $V^{\xi_{\rm phys}}=V_{ne}$, we are
switching continuously from a system of $N$ free electrons
interacting with a modified potential $w^{\xi=0}(r_{12})$,
to the physical system. That is, $f(r_{12})$ is kept
fixed as $\xi$ varies by means of a suitable electron-electron
interaction $W^\xi$ while the one-electron density $n(\rv)$
changes with $\xi$.
Again, by the Hellmann-Feynmann theorem, we find
\begin{eqnarray}
\frac{\partial E_0^{\xi}}{\partial \xi}   =  
\langle \Psi^{\xi} |\frac{\partial W^{\xi}}{\partial \xi}+
\frac{\partial V^{\xi}}{\partial \xi}|\Psi^{\xi}\rangle =
\nonumber \\
= \int f(r_{12})\frac{\partial w^{\xi}(r_{12})}
{\partial \xi}d\rv_{12}+
 \int n^{\xi}(\rv)\frac{\partial v^{\xi}(\rv)}
{\partial \xi}d\rv,
\label{eq_HFAPDFT}
\end{eqnarray}
so that
\beq
G[f;V_{ne},T]=T_{\rm f}[f]+\int_0^{\xi_{\rm phys}} d\xi\int d\rv\,
n^{\xi}(\rv)\frac{\partial v^{\xi}(\rv)}
{\partial \xi},
\label{eq_adiaf}
\eeq
where $T_{\rm f}[f]$ is the kinetic energy of a system of $N$ free
fermions (zero external potential) having the same $f(r_{12})$ of the 
physical 
system. A simple example of such adiabatic connection is given
in Sec.~\ref{sec_harm}. As we shall see, given a confined physical system,
the corresponding $w^{\xi=0}(r_{12})$ must be partially attractive
(in order to create a bound cluster of fermions). This could in principle
lead to ``exotic'' ground states for some of the 
systems corresponding to $\xi<\xi_{\rm phys}$. This issue
is not considered in this paper, and will be investigated in future work.

Similarly to the DFT case, it could be convenient to choose 
as starting point
a system with an external potential corresponding to some coupling
constant $\xi$ (say, $\xi=\alpha$) between 0 and $\xi_{\rm phys}$. If
$\Psi^\alpha$ is the ground-state wavefunction of the system with
external potential $V^\alpha$ (and e-e interaction $W^\alpha$) we have
\begin{eqnarray}
G[f;V_{ne},T]& = & \langle \Psi^{\alpha}|T+V^\alpha|\Psi^{\alpha}\rangle +
\nonumber \\
& & +\int_\alpha^{\xi_{\rm phys}} d\xi\int d\rv\,
n^{\xi}(\rv)\frac{\partial v^{\xi}(\rv)}
{\partial \xi}.
\end{eqnarray}

\subsection{DFT -- Kohn-Sham equations}
One-particle equations in DFT can be obtained by defining
a set of orthogonal orbitals $\varphi_i(\rv)$ with occupation number
$\nu_i$ that minimize 
$\sum_i\nu_i\langle\varphi_i|-\frac{1}{2}\nabla^2|\varphi_i\rangle$
and yield the density of the physical system,
$\sum_i \nu_i|\varphi_i(\rv)|^2=n(\rv)$. This gives
\begin{eqnarray}
\left[-\tfrac{1}{2}\nabla^2+v_1(\rv)\right]\varphi_i(\rv) & = &
\epsilon_i \,\varphi_i(\rv) \nonumber \\
\sum_i \nu_i|\varphi_i(\rv)|^2 & = & n(\rv),
\label{eq_KS}
\end{eqnarray}
where the potential $v_1(\rv)$ is the Lagrange parameter for the density.
To fully specify these equations one needs a rule for the occupation
$\nu_i$ of the orbitals. The Kohn-Sham choice
corresponds to occupy the orbitals in the same way as for
a Slater determinant. 
This determinant is the
wavefunction of a system of $N$ non-interacting
electrons constrained to have the same one-electron density of the
physical system, and leads to  the identification
\beq
T_s[n]=\min_{\{\varphi_i\}\to n}
\sum_i \langle \varphi_i|-\frac{1}{2}\nabla^2|\varphi_i\rangle,
\eeq
with the same $T_s[n]$ of Eq.~(\ref{eq_defTs}).
The ground-state energy of the physical system is then obtained
via the Hartree-exchange-correlation functional $E_{\rm Hxc}[n]$,
defined as the difference $F[n;V_{ee},T]-T_s[n]$. This also 
implies that, in Eqs.~(\ref{eq_KS}),
$v_1(\rv)=v_{\rm KS}(\rv)=v_{ne}(\rv)+\delta E_{\rm Hxc}[n]/\delta n(\rv)$.

Usually, the  Kohn-Sham equations are derived starting from the
 noninteracting system with density $n(\rv)$, rather than from
a constrained minimization of $\sum_i\nu_i\langle \varphi_i|-\frac{1}{2}
\nabla^2|\varphi_i\rangle$. This different way of proceeding allows us
to keep the analogy with what we will do in the next subsection
for APDFT.

%Another possible choice is to occupy only one orbital, which leads
%to an equation  for
%$\sqrt{n(\rv)}$~\cite{PLS}, with a different partitioning of 
%$F[n;V_{ee},T]$ and a different $v(\rv)$ in Eqs.~(\ref{eq_KS}).

\subsection{APDFT -- effective equations}
\label{sec_feffe}
Since the e-e interaction is spherically symmetric, the relevant APD
that determines $\langle\Psi|V_{ee}|\Psi\rangle$ is a unidimensional quantity.
To obtain simple ``two-particle'' equations for $f(r_{12})$ we start from
the kinetic energy operator for the scalar relative coordinate
$r_{12}=|\rv_2-\rv_1|$,
\beq
T_{12}=-\nabla^2_{r_{12}}=-\frac{1}{r_{12}}\frac{d^2}{dr_{12}^2}r_{12}
+\frac{\ell (\ell+1)}{r_{12}^2},
\eeq
and we perform 
a minimization of 
$\sum_i \vartheta_i \langle \psi_i|-\nabla^2_{r_{12}}|\psi_i\rangle$ 
with respect to some orthogonal
``effective'' geminals $\psi_i(r_{12})$ constrained to yield
$f$ of the physical system, $\sum_i \vartheta_i|\psi_i|^2=f$,
leading to
\begin{eqnarray}
& & [-\nabla^2_{r_{12}}+w_{\rm eff}(r_{12})] \psi_i(r_{12})  =  \epsilon_i\,
\psi_i(r_{12}) 
\nonumber \\ 
& & \sum_i \vartheta_i|\psi_i(r_{12})|^2  =  f(r_{12}).
\label{eq_eff}
\end{eqnarray}
The interaction $w_{\rm eff}(r_{12})$ is the Lagrange parameter for $f$. 
Again, to fully specify these equations
we need a rule for the occupancy $\vartheta_i$ 
of the effective geminals. For spin compensated systems, we can choose
to apply a rule that resembles to a Slater determinant: 
occupancy 1 for even $\ell$ (singlet symmetry), occupancy
3 for odd $\ell$ (triplet symmetry), up to $N(N-1)/2$ occupied geminals.
This rule has been applied to solve the effective equations~(\ref{eq_eff})
in the uniform electron gas, with rather accurate 
results~\cite{GP1,DPAT1,CGPenagy2}. It is however important to
point out that when we apply this occupancy rule to 
Eqs.~(\ref{eq_eff})
\begin{enumerate}
\item there is no Slater determinant that can be associated
with our effective geminals: the $\psi_i$ are constrained
to give the {\em exact} $f$ that cannot be obtained from
a noninteracting wavefunction (for example, any Slater determinant
violates the cusp condition satisfied by the exact $f$);
\item more generally, there is no wavefunction (and so no physical
system) that can be built from our effective geminals.
\end{enumerate}
 This last point implies that, if we define
\beq
T_g[f]=\min_{\{\psi_i\}\to f}
\sum_i \langle \psi_i|-\nabla^2_{r_{12}}|\psi_i\rangle
\eeq
(with the determinant-like occupancy), we have in general
\beq
T_g[f] \neq T_{\rm f}[f],
\eeq
where $T_{\rm f}[f]$ was defined in Eq.~(\ref{eq_adiaf}).
The total energy of the physical system can then be recovered via the 
kinetic and external-potential functional defined in Ref.~\cite{GS1},
$F_{\rm KE}[f;V_{ne}]=G[f;V_{ne},T]-T_g[f]$. This also leads to
the identification, in Eqs~(\ref{eq_eff}),
$w_{\rm eff}(r_{12})=1/r_{12}+\delta F_{\rm KE}[f;V_{ne}]/\delta f(r_{12})$.

An important issue to be addressed concerning the
effective equations~(\ref{eq_eff}) is whether a given 
physical (and thus $N$-representable)
$f(r_{12})$ is also representable by the simple 
``effective-geminal'' decomposition of Eqs.~(\ref{eq_eff}). This question
is similar to the one arising in DFT: is a physical
density always non-interacting representable? In view of the
more complex nature of $f(r_{12})$ with respect to $n(\rv)$
we might expect that this problem is much more difficult to face
in APDFT than in DFT. It seems reasonable that at least the
short-range part of a physical $f(r_{12})$ is representable
by Eqs.~(\ref{eq_eff}), while the long-range tail of $f$ of an
extended system could be problematic~\cite{newziesche}.

\section{Simple physical examples: two-electron systems}
\label{sec_ex}
In this section we give some examples for two-electron systems, 
in order to gain
physical insight with some of the ideas just introduced.

\subsection{A picture from harmonic forces}
\label{sec_harm}
A very simple picture of the whole adiabatic connection path in 
both DFT and APDFT
can be gained by looking at an analytic-solvable model, i.e., a
two-electron hamiltonian with only harmonic forces (harmonic electron-nucleus 
attraction, and harmonic e-e repulsion too)~\cite{davidson},
\beq
H(K,k)=-\tfrac{1}{2}\nabla^2_1-\tfrac{1}{2}\nabla^2_2
+\tfrac{1}{2}\,K\,r_1^2+\tfrac{1}{2}\,K\,r_2^2
-\tfrac{1}{2}\,k\,|\rv_1-\rv_2|^2,
\label{eq_Hharmonic}
\eeq
where $K>0$ (attractive nucleus-electron potential), 
$k>0$ (repulsive e-e interaction), and $K>2k$, to have a bound system.
For this hamiltonian,
\begin{eqnarray}
n(r)=\frac{2 \beta^{3/2}}{\pi^{3/2}}\,e^{-\beta\,r^2},
& \;\; & \beta=\frac{2\sqrt{K\,(K-2k)}}{\sqrt{K-2k}+\sqrt{K}} 
\label{eq_nharm}\\
f(r_{12})=\frac{\gamma^{3/2}}{\pi^{3/2}}\,e^{-\gamma\,r_{12}^2},
& \;\; & \gamma=\frac{1}{2}\sqrt{K-2k}.
\label{eq_fharm}
\end{eqnarray}
\subsubsection{DFT}
\begin{figure}
\includegraphics[width=6.5cm]{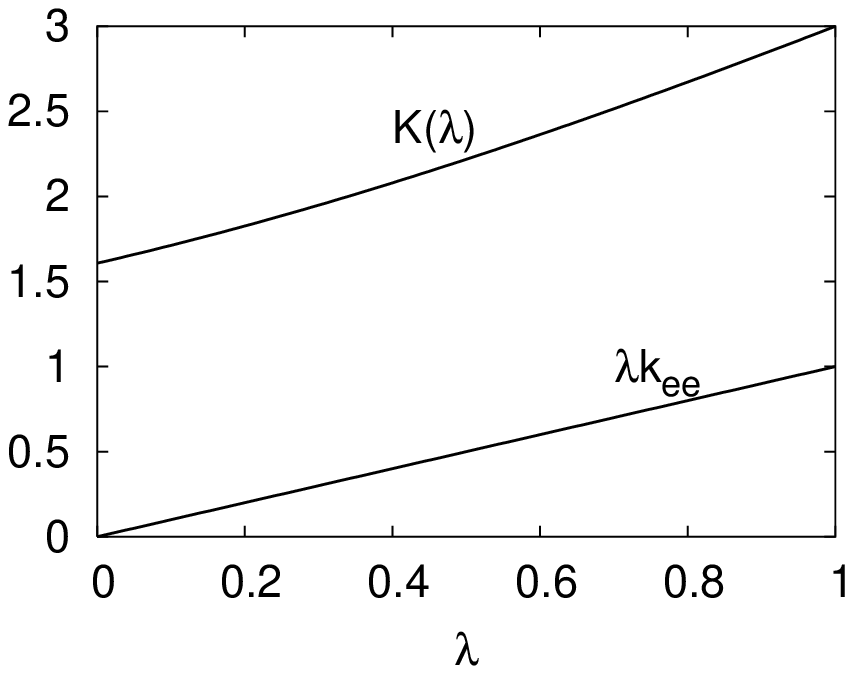} 
\includegraphics[width=6.5cm]{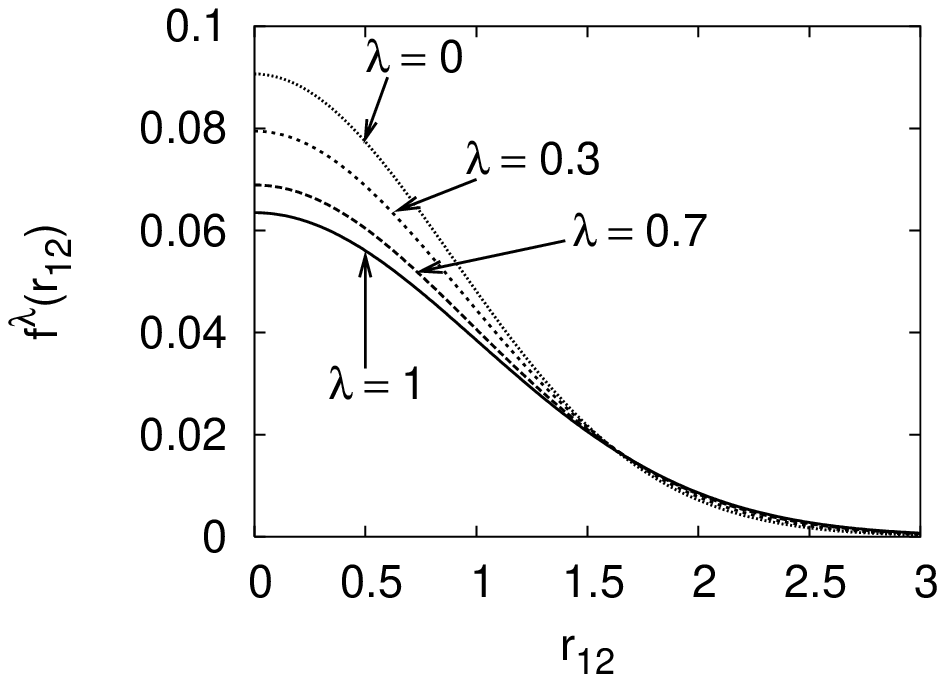} 
\caption{Adiabatic connection in DFT for the simple harmonic hamiltonian
of Eq.~(\ref{eq_Hharmonic}). The e-e interaction is multiplied by
a parameter $\lambda$, and the density is kept fixed by a suitable external
potential (upper panel). 
The APD $f(r_{12})$ changes with $\lambda$ as shown in the
lower panel.}
\label{fig_harm_DFT}
\end{figure}

If our ``physical'' system corresponds to some $K=K_{ne}$ and
$k=k_{ee}$, and we want to switch off the e-e interaction (by
setting, e.g., $k=\lambda k_{ee}$) while keeping the density fixed,
we will simply have to change $K_{ne}$ into $K(\lambda)$ such that
$\beta$ in Eq.~(\ref{eq_nharm}) does not change. 
The function $K(\lambda)$ is shown in the
upper panel of Fig.~\ref{fig_harm_DFT} for the case $K_{ne}=3$, $k_{ee}=1$.
We see that, as $\lambda \to 0$, $K(\lambda)$ decreases, because a smaller 
attraction is needed to keep the electrons in the density when there is no
e-e repulsion. 
Of course,
$f(r_{12})$ changes with $\lambda$, as shown in 
in the lower panel of Fig.~\ref{fig_harm_DFT}: as $\lambda$ 
decreases, the ``on-top'' value $f^{\lambda}(r_{12}=0)$ gets
larger. This reflects the fact that when there is no e-e repulsion, 
it is more likely to find the two electrons close to each other. 
\subsubsection{APDFT}
\begin{figure}
\includegraphics[width=6.5cm]{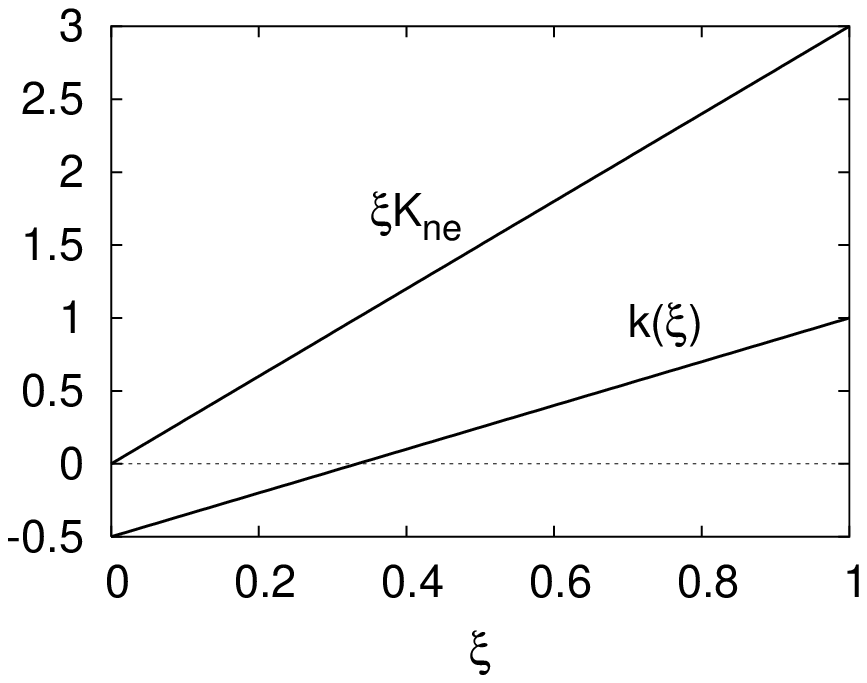} 
\includegraphics[width=6.5cm]{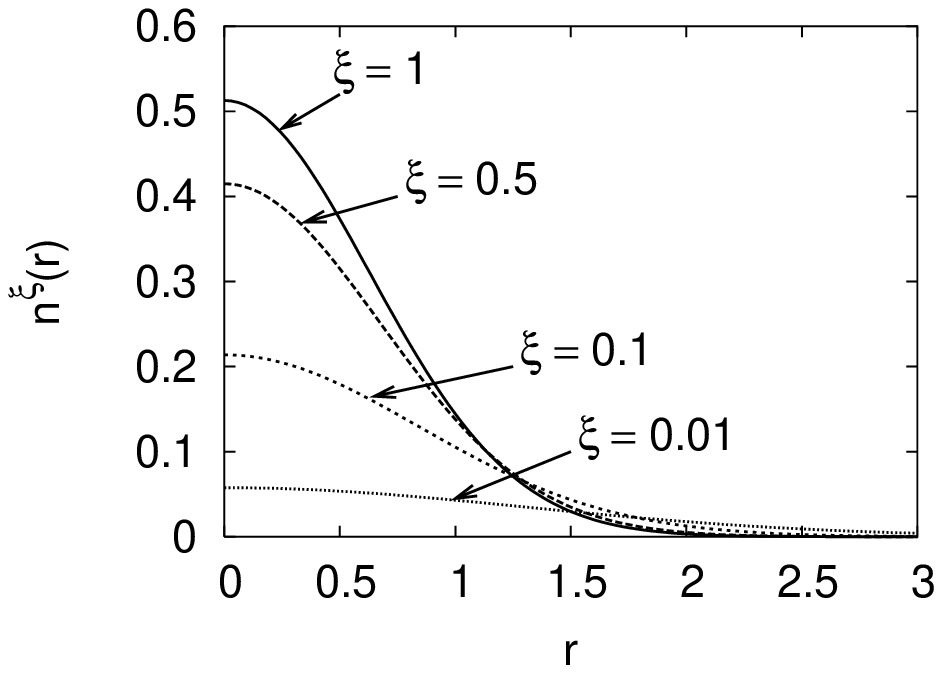} 
\caption{Adiabatic connection in APDFT for the simple harmonic hamiltonian
of Eq.~(\ref{eq_Hharmonic}). The external potential is multiplied by
a parameter $\xi$, and the function $f(r_{12})$ is kept fixed by a suitable 
electron-electron interaction (upper panel). 
The density $n(r)$ changes with $\xi$ as shown in the
lower panel, and as $\xi\to 0$ becomes completely delocalized.}
\label{fig_harm_FFT}
\end{figure}

If, instead, we switch off the external potential, 
(e.g., by setting $K= \xi K_{ne}$)
while keeping $f(r_{12})$ fixed, we will
have to change the e-e interaction in order to keep $\gamma$
of Eq.~(\ref{eq_fharm}) constant,
\beq
k(\xi)   =  \tfrac{1}{2} (\xi-1) K_{ne} + k_{ee}.
\label{eq_kee}
\eeq
The one-electron density along this adiabatic
connection is
\begin{eqnarray}
n^\xi(r) & = & \frac{2 \beta(\xi)^{3/2}}{\pi^{3/2}}\,
e^{-\beta(\xi)\, r^2}, \nonumber \\
 \beta(\xi) & = & \frac{2\sqrt{\xi K_{ne}\,
(K_{ne}-2k_{ee})}}{\sqrt{K_{ne}-2k_{ee}}+\sqrt{\xi K_{ne}}}. 
\end{eqnarray}
Thus $n^{\xi}(r)$ is a gaussian that, as $\xi \to 0$,
becomes more and more spread, as shown in the
lower panel of Fig.~\ref{fig_harm_FFT}. When the
external potential goes to zero, the system becomes translationally
invariant and
the wavefunction for the center-of-mass degree of freedom
is simply a plane wave. 
Correspondingly, the electron-electron
interaction changes with Eq.~(\ref{eq_kee}): we see from
the upper panel of Fig.~\ref{fig_harm_FFT} that, as $\xi$ gets smaller,
$k({\xi})$ becomes smaller (less repulsive), and then it changes sign at some
$0<\xi<1$, becoming an attractive interaction. For a confined system, 
when the external potential approaches zero, an attractive e-e interaction
is needed in order to keep $f(r_{12})$ fixed.
Moreover, in the very
special case of harmonic forces there is a  value
$\xi^*=1-2k_{ee}/K_{ne}\in (0,1)$ for which the e-e effective interaction
is zero everywhere, $w^{\xi^*}=0$.

Of course, when the e-e physical interaction is the Coulomb 
potential $1/r_{12}$ this 
cannot happen: a system with the same $f(r_{12})$
of the physical system cannot have $w=0$ everywhere. This can 
be simply  understood by thinking that there is no external
potential that can force the system to have the correct cusp~\cite{cusp}
at $r_{12}=0$.
In fact, along any adiabatic connection that keeps $f(r_{12})$ fixed,
equal to the one of a system with Coulombic e-e interaction,
$w^\xi(r_{12})$ will always behave as $1/r_{12}$ in the limit
of small $r_{12}$.
%%%%%%%%%%%%%%%%%%%%%%%%%%%%%%%%%%%%%%%%%%%%%%%%%%%%%%%%%%%%%%%%%%%%
%%%%%%%%%%%%%%%%%%%%%%%%%%%%%%%%%%%%%%%%%%%%%%%%
%%%%%%%%%%%%%%%%%%%%%%%%%%%%%%%%%%%%%%%%%%%%%%%%%%%%%%%%%%%%%%%%%%%%
%%%%%%%%%%%%%%%%%%%%%%%%%%%%%%%%%%%%%%%%%%%%%%%%
 
\subsection{He atom}
Consider now the two electrons of a He atom. They feel the attraction of the
nucleus, $-2/r$, and they repel each other with potential
$1/r_{12}$. Given the exact (or a very accurate~\cite{morgan}) 
ground-state wavefunction $\Psi$, we can
calculate the ``exact'' density $n(\rv)$ and the ``exact'' $f(r_{12})$.
Now, we can consider the case in which $W=0$ and $n(\rv)$ is kept
fixed (the KS system in DFT, Sec.~\ref{sec_DFTadia}), 
and the one in which $V=0$ and $f(r_{12})$ is kept 
fixed (APDFT, Sec.~\ref{sec_APDFTadia}):
\subsubsection{DFT}
\begin{figure}
\includegraphics[width=6.5cm]{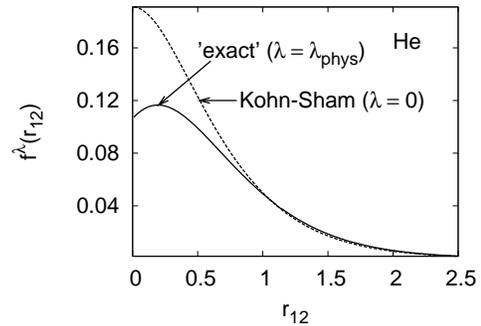} 
\caption{The function $f^{\lambda}(r_{12})$ at the two ends 
of the adiabatic connection in DFT for the He atom. For $\lambda=0$, we have 
a system of two noninteracting ($W=0$) electrons constrained by an external 
potential to yield the same density of the physical system. For
$\lambda=\lambda_{\rm phys}$ we have the physical 
system with full interaction $1/r_{12}$
and external potential $-2/r$ (from the wavefunction of Ref.~\cite{morgan};
see also Ref.~\cite{GS1}).}
\label{fig_comp1}
\end{figure}

We construct a system which has the same
density $n(\rv)$ of the physical one and no electron-electron interaction.
This is the KS system, in which the
two electrons do not interact ($w=0$) and feel an external 
potential $v(r)$ less attractive
than $-2/r$, as in the case of the harmonic potential
of Fig.~\ref{fig_harm_DFT}.
The APD $f^{\lambda=0}(r_{12})$ of this
system will
be different from the physical one, as shown in Fig.~\ref{fig_comp1}.
We see that the change in the function $f$ when we switch from the physical
system to the KS one is qualititatively similar to the one of
Fig.~\ref{fig_harm_DFT}, i.e., at $\lambda=0$ the ``on-top'' value
is higher than the physical one. In the case of Coulomb e-e
interaction the physical $f(r_{12})$ has a cusp, $f'(0)=f(0)$, due
to the short-range divergence of $1/r_{12}$~\cite{cusp}.

\subsubsection{APDFT} 
\begin{figure}
\includegraphics[width=6.5cm]{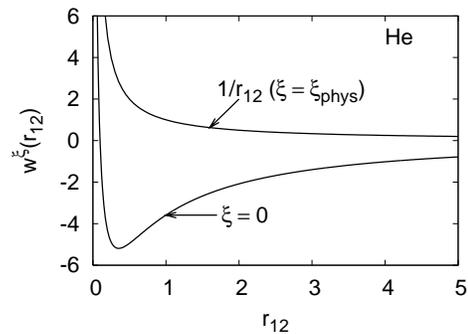} 
\caption{The electron-electron interaction $w^\xi(r_{12})$
at the two ends of the adiabatic connection in APDFT for the He atom. For
$\xi=0$ we have a system of two free fermions (zero external potential)
interacting with $w^{\xi=0}(r_{12})$. This system has the same
$f(r_{12})$ of the physical system, but a completely delocalized one-electron
density. For $\xi=\xi_{\rm phys}$ 
we have the physical system, with e-e interaction
 $1/r_{12}$ and external potential $-2/r$. (The potential $w$ 
at $\xi=0$ has been calculated from the accurate wavefunction of
Ref.~\cite{morgan}; see Ref.~\cite{GS1} for more details.)}
\label{fig_potHe}
\end{figure}

In APDFT, we can construct a system which has the same $f(r_{12})$ of the
physical one, and zero external potential ($V=0$). This is a system
of two bounded fermions interacting with the effective potential 
$w^{\xi=0}(r_{12})$ 
of Fig.~\ref{fig_potHe} (calculated from an accurate~\cite{GS1,morgan}
$f$). As in the case of harmonic forces,
the density of this system is completely delocalized, because
the wavefunction for the
center-of-mass degree of freedom is a plane wave.
We can imagine that along 
the linear adiabatic connection, $v^\xi(r)=-2\xi/r$, the
corresponding $w^\xi(r_{12})$ changes smoothly between
$1/r_{12}$ (at $\xi=\xi_{\rm phys}=1$) and the potential 
$w^{\xi=0}(r_{12})$ of Fig.~\ref{fig_potHe}.
As anticipated, we se that at $\xi=0$ the effective
e-e interaction has an attractive part, which is necessary to
have the same $f$ of a physical confined system.
However, as already pointed out, 
for small $r_{12}$, the e-e interaction must always
behave as $1/r_{12}$, to produce the exact cusp in 
$f(r_{12})$~\cite{cusp}. 

%%%%%%%%%%%%%%%%%%%%%%%%%%%%%%%%%%%%%%%%%%%%%%%%%%%%%%%%%%%%%%%%%%%%
%%%%%%%%%%%%%%%%%%%%%%%%%%%%%%%%%%%%%%%%%%%%%%%%
%%%%%%%%%%%%%%%%%%%%%%%%%%%%%%%%%%%%%%%%%%%%%%%%%%%%%%%%%%%%%%%%%%%%
%%%%%%%%%%%%%%%%%%%%%%%%%%%%%%%%%%%%%%%%%%%%%%%%

\section{From APDFT to correlation energy functionals for DFT}
\label{sec_comb}
In Sec.~\ref{sec_formalism} we have underlined the similarity of the roles
played by $n(\rv)$ and $f(r_{12})$ from a mathematical point of view: the
former completely determines $\langle \Psi |V_{ne}|\Psi\rangle$, and the latter
$\langle \Psi |V_{ee}|\Psi\rangle$. 
However, while the KS system is a substantial 
simplification
of the many-electron problem (yielding to single particle equations), 
the auxiliary system of Sec.~\ref{sec_APDFTadia} 
(with zero external potential) is still a complicated
many-body object, consisting of $N$ fermions interacting
 with a partially attractive potential. The radial equations
of Sec.~\ref{sec_feffe} are a  great simplification of the problem, but
we might expect that building approximations for the whole functional 
$F_{\rm KE}[f;V_{ne}]$ could be not easy. 

Our basic idea, instead, is to use
APDFT to build what is missing in DFT, 
i.e., to build $f^{\lambda}(r_{12})$ along the adiabatic connection in DFT. 
As said in the Introduction, 
we insert our approach in the framework of exact-exchange 
DFT~\cite{FNM,mike,newjohn,oep} in which only the correlation
energy functional needs to be approximated. The ground-state 
energy of the physical system is given by
\beq
E_0=T_s[n]+\int n(\rv) v_{ne}(\rv)d\rv +E_{\rm H}[n]+E_x[n]+E_c[n],
\label{eq_funzionali}
\eeq
where $E_{\rm H}[n]$ is the usual Hartree term, $E_x[n]$ is the
exchange energy, obtained by putting the Kohn-Sham orbitals in the
 Hartree-Fock expression for exchange, and
the correlation energy, $E_c[n]$, is unkown. 
Equation~(\ref{eq_funzionali}) can be also rewritten as
\beq
E_0=\langle \Phi_{\rm KS}|T+V_{ee}+V_{ne}|\Phi_{\rm KS}\rangle+E_c[n],
\label{eq_defEc} 
\eeq
where $\Phi_{\rm KS}$ is the Slater determinant of Kohn-Sham orbitals, i.e.,
the wavefunction of $N$ noninteracting electrons constrained
to yield the same $n(\rv)$ of the physical system.
Combining Eq.~(\ref{eq_defTs}) with Eq.~(\ref{eq_defEc}), we see that
the wanted correlation energy is given by
\beq
E_c[n]=\int_0^{\lambda_{\rm phys}} d\lambda \int_0^\infty dr_{12}\, 
4\pi\,r_{12}^2\,f_c^{\lambda}(r_{12})\frac{\partial w^{\lambda}(r_{12})}
{\partial \lambda},
\label{eq_Ecfromf}
\eeq
where
\beq
f_c^{\lambda}(r_{12})=f^{\lambda}(r_{12})-f^{\lambda=0}(r_{12})
=f^{\lambda}(r_{12})-f_{\rm KS}(r_{12}).
\eeq
Thus, in order to get the KS correlation energy, 
we should compute $f^\lambda(r_{12})$
for each hamiltonian $H^{\lambda}$ of Sec.~\ref{sec_DFTadia}. Our
approach consists in solving the simple radial equations of 
Sec.~\ref{sec_feffe} for each $H^\lambda$ along the adiabatic connection
in DFT. This is not particularly expensive: we are dealing with 
unidimensional equations, and, if the dependence of $w^\lambda$ on
$\lambda$ is smooth, we will only need few $\lambda$ values ($\sim 5-30$)
between 0 and $\lambda_{\rm phys}$. With this particular combination
of APDFT and DFT, we do not need to approximate the whole
functional $F_{\rm KE}$ along the DFT
adiabatic connection, but only its functional derivative, i.e.,
the effective interaction $w_{\rm eff}(r_{12})$ which appears
in Eqs.~(\ref{eq_eff}), since the remaining information is provided
by DFT. As we shall see, simple physical arguments can be used to
build reasonable approximations for $w_{\rm eff}(r_{12})$
at each coupling strength $\lambda$.

Since the effective equations yielding $f_c^\lambda(r_{12})$
must be solved for {\em each} system, we speak of
{\em system-adapted} correlation energy density functionals.

\section{Building approximations}
\label{sec_app}
In Ref.~\cite{GS1}, we proposed and successfully tested
a simple approximation for building
$w_{\rm eff}(r_{12})$ along the DFT adiabatic connection 
for two-electron atoms. This approximation starts from
$w_{\rm eff}^{(0)}(r_{12})$, the effective e-e interaction
that gives $f_{\rm KS}(r_{12})$ when inserted in Eqs.~(\ref{eq_eff}).
In the special case of two-electron systems, $w_{\rm eff}^{(0)}(r_{12})$
is directly available in a KS calculation.
For systems with more than two electrons, $w_{\rm eff}^{(0)}(r_{12})$
 could be calculated, e.g., with the methods of Refs.~\cite{francois,parr}.
Then, the idea is to build an approximation for a  
correlation potential, to be added to $w_{\rm eff}^{(0)}(r_{12})$,
which describes the change in $f$ when the e-e interaction is turned on,
from zero to $w^\lambda(r_{12})$.
%while the density $n(\rv)$ is kept fixed..
%It is intuitive to think that this correlation potential must be
%a screened interaction, with screening lenght related to the ``size''
%of the system.
To do this, we defined an average density $\overline{n}$,
\beq
\overline{n}=\frac{1}{N}\int d\rv\,n(\rv)^2,
\label{eq_avn}
\eeq
and, correspondingly, an average radius $\overline{r}_s$,
\beq
\overline{r}_s=\left(\tfrac{4 \pi}{3}\,\overline{n}\right)^{-1/3}.
\label{eq_rsav}
\eeq  
We then built a correlation potential 
$w^{c,\lambda}_{\rm eff}(r_{12})$, as
\beq
w^{c,\lambda}_{\rm eff}(r_{12}) =w^{\lambda}(r_{12})-
\int_{|\rv|\le \overline{r}_s} 
\overline{n}\,  w^{\lambda}(|\rv - \rv_{12}|)\,d \rv.
\label{eq_vefflambda}
\eeq
%to be added to the one that generates
%$f_{\rm KS}$,
%\beq
%w^\lambda_{\rm eff}(r_{12})\approx w_{\rm eff}^{(0)}(r_{12})+
%w^{c,\lambda}_{\rm eff}(r_{12}).
%\eeq
The idea behind Eq.~(\ref{eq_vefflambda}) is the following:
the e-e interaction $w^\lambda$ is screened by a sphere of radius 
$\overline{r}_s$ and of positive
uniform charge of density $\overline{n}$ that attracts the electrons
with the same interaction $w^\lambda$. The average density 
$\overline{n}$ of Eq.~(\ref{eq_avn})
(and thus the average $\overline{r}_s$) is kept
fixed to mimic the fact that the one-electron density does not change along the
adiabatic connection.
\par

%%%%%%%%%%%%%%%%%%%%%%%%%%%%%%%%%%%%%%%%%%%%%%%%%%%%%%%%%%%%%%%%%%%%
\begin{figure}
\includegraphics[width=6.5cm]{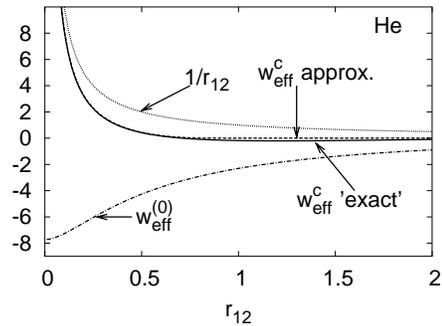} 
\caption{Construction of an approximation for the effective
potential that generates the APD $f(r_{12})$ of the He atom:
$w_{\rm eff}^{(0)}$ is the part of the potential that generates
the APD of the Kohn-Sham system. The ``exact''~\cite{morgan,GS1}
correlation potential $w_{\rm eff}^c$ and our approximation
of Eq.~(\ref{eq_Ov}) are shown, together with the Coulomb
repulsion $1/r_{12}$.}
\label{fig_OvHe}
\end{figure}
%%%%%%%%%%%%%%%%%%%%%%%%%%%%%%%%%%%%%%%%%%%%%%%%

In order to gain insight with our construction, let us consider the case of
the physical system, $w^{\lambda=\lambda_{\rm phys}} =1/r_{12}$, 
for which Eq.~(\ref{eq_vefflambda}) corresponds to
\beq
w^c_{\rm eff}(r_{12})=\left(\frac{1}{r_{12}}+
\frac{r_{12}^2}{2\,\overline{r}^3_s}-\frac{3}{2\,\overline{r}_s}\right)
\theta\left(\overline{r}_s-r_{12}\right),
\label{eq_Ov}
\eeq   
where $\theta(x)$ is the Heaviside step function. In Fig.~\ref{fig_OvHe}
we reported, for the He atom, the potential $w_{\rm eff}^{(0)}$
which generates $f_{\rm KS}$, together with the ``exact'' correlation
potential $w^c_{\rm eff}$, and the approximation of Eq.~(\ref{eq_Ov}).
We see that the potential $w_{\rm eff}^{(0)}$ is a confining
potential for the variable $r_{12}$: our idea is to include in this
term, available from DFT, the contribution to
$f(r_{12})$ coming from the particular external potential of
the system and from the fermionic structure of the wavefunction.
The remaining part to be approximated, the correlation
potential $w_{\rm eff}^c$, must include the effect of the e-e repulsion
while keeping the density fixed, i.e., it must be
essentially  a screened Coulomb 
interaction. We see from Fig.~\ref{fig_OvHe} that the simple approximation
of Eq.~(\ref{eq_Ov}) is reasonable, i.e., the screening
length is well approximated by $\overline{r}_s$ of 
Eqs.~(\ref{eq_avn})-(\ref{eq_rsav}). For comparison, the full
Coulomb repulsion $1/r_{12}$ is also shown.
Notice that in the special case of two-electron systems we have
$T_g[f]=T_{\rm f}[f]$, so that the potential $w^{\xi=0}$
of Fig.~\ref{fig_potHe} corresponds, in  Fig.~\ref{fig_OvHe}, to
the sum of $w_{\rm eff}^{(0)}$ and the ``exact'' $w_{\rm eff}^c$.

In Ref.~\cite{GS1}, we inserted the potential $w_{\rm eff}^{(0)}(r_{12})+
w^{c,\lambda}_{\rm eff}(r_{12})$ into Eqs.~(\ref{eq_eff}),
and solved them for several two-electron atoms. Our results can
be summarized as follows:
(i) at $\lambda=\lambda_{\rm phys}$
(i.e., for $w^\lambda(r_{12})=1/r_{12}$) we obtained APD $f(r_{12})$
in close agreement with those coming from accurate
variational wavefunctions~\cite{morgan}, especially
at small $r_{12}$; (ii) by setting $w^\lambda(r_{12})=
\erf(\lambda r_{12})/r_{12}$ (the ``erf'' adiabatic
connection), the KS correlation energies 
from Eq.~(\ref{eq_Ecfromf}) have an
error which is less than 4~mH for nuclear charges
$Z\ge 2$; (iii) again with the ``erf''
adiabatic connection, we found that 
when the reference system corresponds to some $\lambda=\mu$
between zero and $\lambda_{\rm phys}$ [as in Eq.~(\ref{eq_Fmu})]
our correlation energies
are one order of magnitude better for $\mu \gtrsim 1/\overline{r}_s$.

The correlation potential of Eq.~(\ref{eq_Ov}),
originally proposed by Overhauser~\cite{Ov}, has been also used
to solve the effective Eqs.~(\ref{eq_eff}) for the uniform electron gas
(UEG), yielding to a very accurate description of the
short-range part of $f(r_{12})$ at all densities~\cite{GP1}.
A more sophisticated effective potential, based on 
a self-consistent Hartree
approximation, extended such accuracy to the long-range part
of the UEG $f(r_{12})$ at metallic densities~\cite{DPAT1}.
Other simple approximations for $w_{\rm eff}(r_{12})$ in the UEG
have also been proposed and tested~\cite{CGPenagy2}.

\section{Conclusions and perspectives}
\label{sec_last}
We have presented the ideas concerning a theory based on the
spherically and system-averaged pair density $f(r_{12})$, and
we have suggested to combine it with DFT to obtain system-adapted
correlation energy functionals. So far, the method has been
tested for the He series~\cite{GS1} and for the uniform electron 
gas~\cite{GP1,DPAT1,CGPenagy2}, yielding promising results.
 
In order to completely develop the approach presented here, many
steps are still to be performed and many issues are to be addressed.
Among them, the most relevant ones concern the construction of better
approximations for the effective electron-electron interaction that
enters the formalism, and the implementation of a self-consistent scheme
to combine the Kohn-Sham equations with the correlation energy functional
arising from our approach. Last but not least, with the approximations
tested so far our approach works very well for the short-range part
of $f(r_{12})$, so that the combination with multideterminantal
DFT~\cite{adiabatic,erf} (in which only the short-range correlations
are treated within DFT) is also very promising and deserves further 
investigation.

\section*{Acknowledgments}
We thank C. Umrigar for the wavefunctions of Ref.~\cite{morgan}.
This research was supported by a Marie Curie Intra-European
Fellowships within the 6th European Community Framework 
Programme (contract number MEIF-CT-2003-500026).

\end{document}